\documentclass[twocolumn,aps,prl,amsmath,amssymb]{revtex4}
\usepackage{graphicx,dcolumn,bm,color,framed}
\begin{document}

\title{Photon-Induced Suppression of Interlayer Tunneling in Van Der Waals Heterostructures}

\author{Woo-Ram Lee}
\author{Wang-Kong Tse}
\affiliation{
Department of Physics and Astronomy, The University of Alabama, Alabama 35487, USA\\
Center for Materials for Information Technology, The University of Alabama, Alabama 35401, USA}

%\date{\today}

\begin{abstract}
We develop a theory for interlayer tunneling in van der Waals heterostructures driven under a strong electromagnetic field, using graphene/{\it h}-BN/graphene as a paradigmatic example. Our theory predicts that strong anti-resonances appear at bias voltage values equal to an integer multiple of the light frequency. These features are found to originate from photon-assisted resonant tunneling transitions between Floquet sidebands of different graphene layers, and are unique to two-band systems due to the interplay of both  intraband and interband tunneling transitions. Our results point to the possibility of tunneling localization in van der Waals heterostructures using strong electromagnetic fields. 

%Our results point to the possibility of  photon-assisted localization of tunneling in van der Waals heterostructures at strong electromagnetic fields. 
%
% We present a theory for electron tunneling in a graphene/{\it h}-BN/graphene van der Waals heterostructure under strong optical illumination. Using the Keldysh-Floquet Green's function formalism, we have obtained the tunneling current between the graphene layers when both layers are driven periodically by an optical field. Our theory predicts that photon-assisted anti-resonance peaks are periodically developed in the tunneling current-voltage characteristics on top of the contribution due to the dark tunneling current. We find that, while the latter
% originates from interband tunneling processes, the anti-resonance peaks originate from intraband tunneling processes between the Floquet subbands in different graphene layers.
\end{abstract}

\maketitle

%%%%%%%%%%%%%%%%%%%%%%%%%%%
%\section{Introduction}
%%%%%%%%%%%%%%%%%%%%%%%%%%%

%\textit{Introduction} --- 
When light is incident on a tunneling junction, inelastic tunneling transitions can occur via the exchange of photons between electrons and the electromagnetic field. This photon-assisted tunneling (PAT) phenomenon was first predicted in the classic work by Tien and Gordon ~\cite{Tien63} in superconductor-insulator-superconductor tunneling junctions \cite{SIS_exp}. PAT has since been studied and observed in many systems \cite{PAT_Rev1}, including semiconductor quantum dots \cite{PAT_QD1,PAT_QD2,PAT_QD3}, double quantum wells \cite{PAT_QW} and superlattices \cite{PAT_SL,PAT_SL2}, and optical lattices \cite{PAT_OL,PAT_OL2}. 
In particular, tunneling dynamics can be suppressed \cite{DL_2,DL_3,DL_4} when the light coupling parameter, given by the ratio of the driving field amplitude and frequency, matches a zero of the Bessel function, a celebrated phenomenon called dynamic localization \cite{DL}. 
%The celebrated phenomenon of dynamic localization \cite{DL} pertains to a suppression of the tunneling dynamics when the light coupling parameter, given by the ratio of the %driving field amplitude and frequency, matches a zero of the Bessel function. 
%frequency and amplitude of incident light are related to the zeros of the Bessel function. 

Van der Waals heterostructures (vdWHs)~\cite{Antonova16} are an emerging class of nanoscale materials 
%that has received much attention recently. 
%Because they can be synthesized at the nanoscale from the bottom up, vdWHs 
that hold great promise as a platform for realizing unconventional electronic properties and desirable functionalities. 
%discovering unconventional electronic properties and designing desirable functionalities. 
%for potential optoelectronic applications. 
Vertically stacked vdWHs exhibit many distinctive properties not available in conventional semiconductor quantum well systems, including enhanced longitudinal and Hall Coulomb drag~\cite{Drag}, tunable metal-insulator transition~\cite{MIT} and extraordinary photovoltaic response~\cite{Geim}. In addition to in-plane transport, vertical tunneling transport in a field-effect tunneling junction  geometry exhibits superior current-voltage characteristics \cite{TunExp1,TunExp2,TunExp3,TunExp4,TunExp5}.

A strong electromagnetic field can provide a heretofore unexplored degree of freedom for tuning the tunneling dynamics in vdWHs. 
%However, there has been little work addressing the PAT effects in vdWHs 
%remain largely unexplored. 
%inducing changes in electronic properties and hence controling the interlayer transport properties in vdWHs. 
%To date, PAT effects in vdWHs remain largely unexplored. 
In this Letter, we theoretically investigate interlayer tunneling in optically driven vertical vdWHs, using graphene/{\it h}-BN/graphene as an archetypical system. Using the Keldysh-Floquet Green's function formalism, we formulate a theory for the non-equilibrium PAT current and elucidate the non-perturbative effects of the driving field on the intraband and interband tunneling transitions. 
%show that multiphoton excitations can open up new channels for both intraband and interband tunneling transitions. 
Our theory predicts a new type of tunneling localization effect where photon-enabled resonant tunneling processes induce a dramatic suppression of the interlayer tunneling current as a function of the bias voltage, precisely at integer multiples of the photon energy. 
%
%that is fundamentally different from the dynamic localization, 
%that photon-enabled 
%where resonant intraband tunneling processes induce a dramatic suppression of the interlayer tunneling current as a function of the bias voltage, precisely at integer multiples of the photon energy. 

%This predicted suppression is of a different nature from the dynamic localization. 
% A strong electromagnetic field can provide an extra degree of freedom for tuning the electronic properties and controling the interband tunneling process in vdWHs. Importantly, we find that multiphoton excitations open up new channels for resonant intraband tunneling that are absent without light. This resonant tunneling process induces a dramatic suppression of the interlayer tunneling current as a function of the bias voltage, precisely at integer multiples of the photon energy. 

%Our studies reveal photon-induced resonances in the tunneling current. 

%reveal a new phenomenon in vertical vdWHs when they are illuminated by a strong optical field: photon-assisted tunneling resonance. Optical field enables further control beyond electrical gating as a new means to achieve 2D nanoelectronics. 

%%%%%%%%%%%%%%%%%%%%%%%%%%%
%\section{Model}
%\label{Section_Model}
%%%%%%%%%%%%%%%%%%%%%%%%%%%

\textit{Model} --- Our tunneling structure consists of two parallel graphene layers, labeled as top ($T$) and bottom ($B$), that are separated by a middle insulating monolayer of hexagonal boron nitride ({\it h}-BN). The layers are perfectly aligned and stacked in the ABA (Bernal) configuration, and a bias voltage $V$ is applied across the top and bottom layers. The low-energy excitations around the $K$ and $K'$ points (labeled by $\xi = \pm 1$, respectively) in the Brillouin zone of each graphene layer is governed by the 2D massive Dirac model up to an energy cutoff $\mathcal{E}_c \approx 3\,\mathrm{eV}$:
%$h_{{\bf k}\xi T,B} = v\hbar (\xi k_x \sigma_{x} + k_y \sigma_{y}) + \frac{\Delta}{2} \sigma_z \pm \frac{e V}{2} \mathbb{I}_2$
\begin{equation}
h_{{\bf k}\xi T,B}
= v\hbar (\xi k_x \sigma_{x} + k_y \sigma_{y}) + {\Delta} \sigma_z/2 \pm {e V} \mathbb{I}_2/2,
\label{SingleLayerGrapheneHamiltonian}
\end{equation}
where $v$ is the Dirac velocity, $\Delta$ is the band gap induced by the {\it h}-BN layer, $-e$ is the electron charge, and $\{\mathbb{I}_2, \sigma_x, \sigma_y, \sigma_z\}$ denote the identity and Pauli matrices in the sublattice-pseudospin (\textit{i.e.}, $a$ and $b$ sites) space. We construct the Hamiltonian of the trilayer system using a  nearest-neighbor hopping approximation, including the coupling between each graphene layer and the {\it h}-BN layer and ignoring the negligible direct hopping between the graphene sheets. Due to its large band gap $\Delta_{\rm BN} \sim 5\,\mathrm{eV}$, we can trace out the {\it h}-BN layer and obtain an effective double-layer Hamiltonian \cite{Comment1} $\tilde{\mathcal{H}} = \sum_{\bf k} \sum_{\xi\in\{+,-\}} \tilde{\psi}_{{\bf k}\xi}^\dag (\tilde{h}_{{\bf k}\xi} + \tilde{\mathcal{W}})\tilde{\psi}_{{\bf k}\xi}$, where $\tilde{\psi}_{{\bf k}\xi}^\dag = (\phi_{{\bf k}\xi T}^\dag, \phi_{{\bf k}\xi B}^\dag)$. Tilde symbolizes the layer-pseudospin (\textit{i.e.}, $T$ and $B$) space. It is convenient to define $\mathbb{I}_\pm = (\mathbb{I}_2 \pm \sigma_{z})/2$, $\tilde{\mathbb{I}}_\pm = (\tilde{\mathbb{I}}_2 \pm \tilde{\sigma}_{z})/2$, $\tilde{\sigma}_\pm = (\tilde{\sigma}_{x} \pm i \tilde{\sigma}_{y})/2$ and use the new basis $\{\tilde{\mathbb{I}}_{+}, \tilde{\mathbb{I}}_{-}, \tilde{\sigma}_{+}, \tilde{\sigma}_{-}\}$ for the layer pseudospins to write  the unperturbed Hamiltonian as $\tilde{h}_{{\bf k}\xi} = h_{{\bf k}\xi T} \otimes \tilde{\mathbb{I}}_{+} + h_{{\bf k}\xi B} \otimes \tilde{\mathbb{I}}_{-}$ and  the interlayer tunneling Hamiltonian as $\tilde{\mathcal{W}} = \mathcal{W}_{TB} \otimes \tilde{\sigma}_{+} + \mathcal{W}_{BT} \otimes \tilde{\sigma}_{-}$. The tunneling matrix elements are given by $\mathcal{W}_{TB} = \mathcal{W}_{BT}^\dag = \mathcal{W}_0 \mathbb{I}_{-}$ with $\mathcal{W}_0 = - 2\Gamma^2 \Delta_{\rm BN}^{-1}$, where $\Gamma$ is the interlayer hopping energy from $a$ site in graphene to $b$ site in \textit{h}-BN (or vice versa) and $\Delta_{\rm BN}$ is the \textit{h}-BN band gap.

%between an a site and a b site 
%$\mathcal{W}_0 = - 2\Gamma_1 \Gamma_2 \Delta_{\rm BN}^{-1}$. 

We consider the two graphene layers to be coupled, for simplicity, to the same optical field by imagining two independent but identical laser sources setup symmetrically on both sides of the vdWH, illuminating the two graphene layers at normal incidence. Thus, the surface electric fields on both layers will be in phase with the same amplitude and frequency. 
%Proposed experimental setups are discussed at the end of the paper.
%At the end of this paper, we discuss an alternative strategy to realize this scenario. 
%Next, regarding optical setup, we symmetrically set up two independent optical sources outside the system to illuminate both graphene layers at the normal incidence in opposite directions. Then, the surface electric fields on both layers are in phase with the same amplitude and frequency. 
Choosing the propagation direction along the $z$ axis, the incident light with electric field amplitude $E$, frequency $\Omega$, and polarization $\vartheta$ is described by the vector potential $\mathbf{A}(t) = (cE/\Omega) [\sin(\Omega t) \hat{x} + \sin(\Omega t + \vartheta) \hat{y}]$. %\cite{Remark_Switching}. 
%Here, a switching protocol is encoded in the function $\zeta(t) = e^{t / \tau_s} \Theta(-t) + \Theta(t)$, with $\tau_s$ being a switch-on time scale and $\Theta(x)$ the step %function, which satisfies $\zeta \rightarrow 0$ for an equilibrium state at $t\rightarrow-\infty$ and $\zeta = 1$ for a NESS
%at $t\geq0$~\cite{Lee17}. 
%In the semiclassical treatment of the optical field, 
The Peierls substitution $\hbar {\bf k} \rightarrow \hbar {\bf k} + {e} {\bf A}(t)/c$ in the original Hamiltonian produces the time-dependent Hamiltonian $\tilde{\mathcal{H}}_{{\bf k}\xi}(t) = \tilde{\mathcal{H}}_{{\bf k}\xi} + \tilde{\mathcal{V}}_{\xi}(t)$, with the interaction $\tilde{\mathcal{V}}_{\xi}(t) = \mathcal{V}_{\xi}(t) \otimes \tilde{\mathbb{I}}_2$, where $\mathcal{V}_{\xi}(t) = \mathcal{V}_0 [\xi \sin(\Omega t) \sigma_{x} + \sin(\Omega t + \vartheta) \sigma_{y}]$.
% \begin{align}
% \mathcal{V}_{\xi}(t)
% & = \mathcal{V}_0 [\xi \sin(\Omega t) \sigma_{x} + \sin(\Omega t + \vartheta) \sigma_{y}].
% %\mathcal{V}_{\xi}(t)
% %& = \mathcal{V}_0 \zeta(t) [\xi \sin(\Omega t) \sigma_{x} + \sin(\Omega t + \vartheta) \sigma_{y}].
% \end{align}
%Here, 
The ratio $\mathcal{V}_0/\Delta = (E/E_0) / (\hbar\Omega/\Delta)$ with $E_0 = \Delta^2/(ev\hbar)$ describes the amplitude of the interaction in dimensionless form.

\textit{Theory} --- Our formalism is developed by treating the interlayer tunneling Hamiltonian as a perturbation while taking into account the optical field non-perturbatively. 
%We start with a generic form of the electric charge operator on layer $\alpha \in \{T,B\}$, $\hat{Q}_\alpha(t) = - e \sum_{\bf k} \sum_{\xi\in\{+,-\}} [a_{{\bf k}\xi\alpha}^\dag(t) %a_{{\bf k}\xi\alpha}(t) + b_{{\bf k}\xi\alpha}^\dag(t) b_{{\bf k}\xi\alpha}(t)]$. 
Noting that the total charge is conserved in the system, the interlayer electric current density takes the form $J(t) = 2 S^{-1} \langle \hat{I}(t) \rangle$, where $2$ counts the spin degeneracy, $S$ is the normalization area, $\langle \cdots \rangle$ is the canonical ensemble average, and $\hat{I}(t) = \partial_t \hat{Q}_{T}(t) = - \partial_t \hat{Q}_{B}(t) = (i/\hbar) [\tilde{\mathcal{H}}(t), \hat{Q}_{T}(t)]$ is the electric current with $\hat{Q}_{T,B}(t)$ being the electric charge operator on layer T and B. 
%$\langle \mathcal{M} \rangle = {\rm Tr}(\rho_0 \mathcal{M})$ is the canonical ensemble average of the operator $\mathcal{M}$ in terms of the initial density matrix $\rho_0 = %e^{-\tilde{\mathcal{H}}/k_B T} / {\rm Tr}(e^{-\tilde{\mathcal{H}}/k_B T})$. 
The time evolution of $J(t)$ is determined from the equal-time lesser Green's function, $G_{{\bf k}\xi}^<(t,t)$. 
%expressed through the lesser Green's function, $[\hbar G_{{\bf k}\xi}^<(t,t)]_{cd;TB} = i \langle d_{{\bf k}\xi B}(t) c_{{\bf k}\xi T}(t) \rangle$ with $c,d \in \{a,b\}$. 
In a non-equilibrium steady state (NESS), the system respects time translational symmetry %, i.e., $\tilde{\mathcal{H}}(t) = \tilde{\mathcal{H}}(t+\tau)$ with periodicity $\tau (= 2\pi/\Omega)$, 
and is thus governed by the Floquet theorem~\cite{Grifoni98}. By using the Floquet mode expansion of the Green's function \cite{Oka_Review}, 
%$[\hbar G_{{\bf k}\xi}^<(t,t)]_{cd;TB} = \sum_{m,n\in\mathbb{Z}} e^{-i(m-n)\Omega t} \int_{-\hbar\Omega/2}^{\hbar\Omega/2} d\hbar\omega [\hat{G}_{{\bf k}\xi}^<(\omega)]_{cd;TB;mn} / (2\pi),~$ 
we obtain the time-averaged interlayer electric current density $J  = - 4 ({e}/{\hbar}) ({1}/{S}) \sum_{\bf k} \sum_{\xi\in\{+,-\}} \int_{-\hbar\Omega/2}^{\hbar\Omega/2} ({d\hbar\omega}/{2\pi})$ $ {\rm Re} {\rm Tr} \big\{ [\mathcal{W}_{BT} \otimes \hat{\mathbb{I}}_{\infty}] [\hat{G}_{{\bf k}\xi}^<(\omega)]_{TB} \big\}$,
where an overhat refers to a quantity in the Floquet space, the subscript ``${\rm TB}$" of the Green's function refers to its off-diagonal element in the layer subspace,  
%connecting the T and B layers, 
``${\rm Tr}$" is the trace over sublattice pseudospins and Floquet modes, and $\hat{\mathbb{I}}_{n}$ is the $n \times n$ identity matrix in the Floquet space.

The lesser Green's function $\hat{G}_{{\bf k}\xi}^{<}$ can be calculated within the Keldysh-Floquet Green's function formalism~\cite{Lee17, Lee14, Tsuji09}. %Under this condition, 
The full, tunneling-coupled Green's function is uniquely determined by the uncoupled Green's function of each layer and the interlayer tunneling Hamiltonian 
%, Eq.~(\ref{InterlayertunnelingHamiltonian}), 
via the Dyson equation in the Keldysh-Floquet space \cite{InCond}. Expansion of $\hat{G}_{{\bf k}\xi}^{<}$ up to first order in $\mathcal{W}_0$ yields the following {\it photon-assisted} tunneling current formula: 
\begin{align}
J_{\rm tun}
& = - 4 \frac{e}{\hbar} \frac{1}{S} \sum_{\bf k} \sum_{\xi\in\{+,-\}} \int_{-\hbar\Omega/2}^{\hbar\Omega/2} \frac{d\hbar\omega}{2\pi} 
\nonumber\\
& ~~~ \times {\rm Re}\left\{ {\rm Tr} \big[ \hat{\mathbb{W}}^\dag_{{\bf k}\xi TB}(\omega) \hat{\mathcal{G}}_{{\bf k}\xi T}^{R}(\omega) \hat{\mathbb{W}}_{{\bf k}\xi TB}(\omega) \hat{g}_{{\bf k}\xi B}^{<}(\omega) \right.
\nonumber\\
&\left. ~~~ + \hat{\mathbb{W}}^\dag_{{\bf k}\xi BT}(\omega) \hat{\mathcal{G}}_{{\bf k}\xi B}^{A}(\omega) \hat{\mathbb{W}}_{{\bf k}\xi BT}(\omega) \hat{g}_{{\bf k}\xi T}^{<}(\omega) \big]\right\},
\label{TunnelingFormula_2}
\end{align}
where $\hat{\mathcal{G}}_{{\bf k}\xi T, B}^{R}(\omega)$ is the retarded Floquet Green's function of the T and B layers, $\hat{g}_{{\bf k}\xi T}^{<}(\omega)$ is the lesser Floquet Green's function in the absence of light. Coupling to the metallic leads provides the energy relaxation mechanism and is captured by a band broadening parameter $\Gamma$ in the Green's functions \cite{InCond}. We have defined an effective {\it photon-dressed} interlayer tunneling Hamiltonian
\begin{align}
\hat{\mathbb{W}}_{{\bf k}\xi\alpha\bar{\alpha}}(\omega) = \mathcal{W}_{\alpha\bar{\alpha}} \big[ \mathbb{I}_2 \otimes \hat{\mathbb{I}}_\infty + \hat{\mathcal{G}}_{{\bf k}\xi\bar{\alpha}}^{R}(\omega) \hat{\mathcal{V}}_\xi \big], 
\label{PhotonDressedInterlayerTunnelingHamiltonian}
\end{align}
where $\alpha \in \{T, B\}$ and the overbar denotes a complement, \textit{e.g.}, $\bar{T} = B$. Eq.~(\ref{TunnelingFormula_2}) together with Eq.~(\ref{PhotonDressedInterlayerTunnelingHamiltonian}) is a central result of this Letter. It shows that non-perturbative photon dressing effects are contained not only in the Floquet Green's function  $\hat{\mathcal{G}}^{R}$, but also in the tunneling Hamiltonian $\hat{\mathbb{W}}$ through Eq.~\eqref{PhotonDressedInterlayerTunnelingHamiltonian}, in which the second term within the brackets represents the modification of the tunneling amplitude by a strong optical field.

\begin{figure}[t]
\centering
\includegraphics[width=0.47\textwidth]
{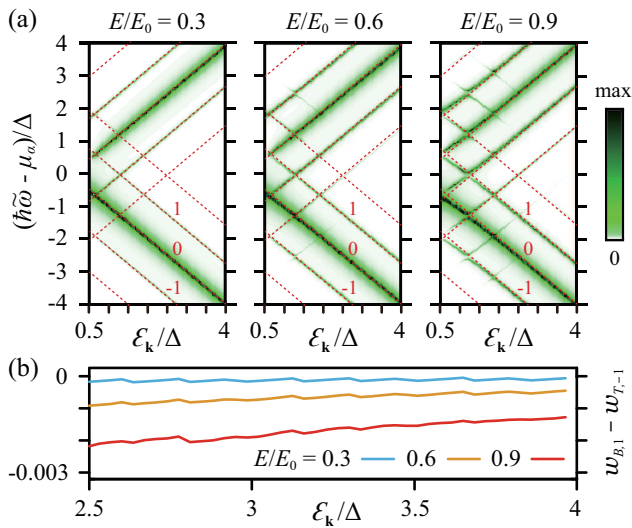} \\
\caption{(a) %Photon-dressed 
Floquet mode spectral function $\rho_{{\bf k}\xi\alpha b}^{\rm (ph)}(\tilde{\omega})$ %[Eq.~\eqref{PhotonDressedSpectralFunction}] 
for valley $K'$, layer $\alpha\in\{T,B\}$, and sublattice $b$ as a function of $\mathcal{E}_{\bf k}$ for different values of $E/E_0$. The red dotted lines are guidelines denoting the equilibrium bands and their corresponding copies shifted by integer multiples of $\hbar \Omega$. 
%means the guideline for the ideal case without the interference between Floquet sidebands. 
Three main Floquet sidebands belonging to $\gamma = -1$ are indexed by red-colored integers. 
%For $E/E_0 = 0.9$, there is a clear deviation from the guideline, indicating the formation of hybridized Floquet states. 
%Here, the common parameters are set by 
Parameters used are $\Gamma / \Delta = 0.01$, $\varphi_{\bf k} = 0$, $\hbar\Omega/\Delta = 1.25$, and $\vartheta = \pi/2$.
(b) Difference between the spectral weights $w_{B,n=1} - w_{T,n=-1}$ of the Floquet valence sidebands $n = \pm 1$ 
%weight factors, $w_{B,n=1}^{-} - w_{T,n=-1}^{-}$, 
as a function of $\mathcal{E}_{\bf k}$ for different values of $E/E_0$. 
%Here, we focus on the range $2.5 \leq \mathcal{E}_{\bf k}/\Delta \leq 4$.
}
\label{Figure1}
\end{figure}

To illustrate the physics contained in Eq.~(\ref{TunnelingFormula_2}) and also to connect with the more familiar case without light, %we 
let us first consider the weak coupling regime \cite{Lee17,weak_coupling} defined as $\mathcal{V}_0 / \hbar\Omega \ll 1$. In this scenario the photon-induced correction $\sim \hat{\mathcal{G}}^{R}\hat{\mathcal{V}}$ to the tunneling Hamiltonian [Eq.~(\ref{PhotonDressedInterlayerTunnelingHamiltonian})] is negligible.   
%When the driving field amplitude is small or its frequency is large such that $\mathcal{V}_0 / \hbar\Omega \ll 1$, the electromagnetic field is weakly coupled to the electrons %and this photon-induced correction will be small. 
%It is negligible in the weak-coupling limit of $\mathcal{V}_0 / \hbar\Omega \ll 1$. 
%In this case, 
For the ABA stacking configuration of our system, $[\mathcal{W}_{TB}]_{cc'} = [\mathcal{W}_{BT}]^*_{c'c} = \mathcal{W}_0 \delta_{cc'} \delta_{cb}$ where $c,c' \in \{a,b\}$ label the sublattice pseudospins, we can reduce Eq.~\eqref{TunnelingFormula_2} to a compact form 
\begin{align}
& \frac{J_{\rm tun}}{J_0}
= \frac{(2\pi v\hbar)^2}{\Delta} \frac{1}{S} \sum_{\bf k} \sum_{\xi \in \{+,-\}} \int_{-\infty}^{\infty} d\hbar\tilde{\omega}
\label{TunnelingFormula_3}
\\
& ~~ \times \big[ \rho_{{\bf k}\xi B b}^{({\rm ph})}(\tilde{\omega}) \rho_{{\bf k}T b}^{({\rm eq})}(\tilde{\omega}) f_{T}(\tilde{\omega}) 
- \rho_{{\bf k}\xi T b}^{({\rm ph})}(\tilde{\omega}) \rho_{{\bf k}B b}^{({\rm eq})}(\tilde{\omega}) f_{B}(\tilde{\omega}) \big], 
\nonumber
\end{align}
where $J_0 = \pi^{-1} e v^{-2} \hbar^{-3} \Delta |\mathcal{W}_0|^2$ and $\tilde{\omega} \equiv \omega + n\Omega$ is the frequency in the extended zone scheme \cite{Oka_Review}. The two terms of Eq.~(\ref{TunnelingFormula_3}) correspond to a forward and a backward tunneling channel, and 
% Eq.~(\ref{TunnelingFormula_3}) elucidates that the photon-dressed tunneling current consists of a forward and a backward tunneling channel.  
%According to this formula, the tunneling current is triggered by breaking the balance between two counter-flowing channels. 
each channel is modified by non-perturbative light coupling effects through the Floquet mode 
%time-averaged {\it photon-dressed} 
spectral function $\rho_{{\bf k}\xi\alpha b}^{\rm (ph)}(\tilde{\omega}) = - ({1}/{\pi}){\rm Im} [\hat{\mathcal{G}}^{R}_{{\bf k}\xi\alpha}(\omega)]_{bb;nn}$. 
%for valley $\xi$, layer $\alpha$, and sublattice $b$
% \begin{align}
% \rho_{{\bf k}\xi\alpha b}^{\rm (ph)}(\tilde{\omega})
% = - \frac{1}{\pi} {\rm Im} [\hat{\mathcal{G}}^{R}_{{\bf k}\xi\alpha}(\omega)]_{bb;nn},
% \label{PhotonDressedSpectralFunction}
% \end{align}
%where the reduced-zone $\omega$ was switched to the extended-zone $\tilde{\omega} (\equiv \omega + n\Omega)$~\cite{Comment2}. 
In the absence of optical fields, %Eq.~\eqref{PhotonDressedSpectralFunction} 
$\rho_{{\bf k}\xi\alpha b}^{\rm (ph)}(\tilde{\omega})$ 
recovers its equilibrium counterpart $\rho_{{\bf k}\alpha b}^{\rm (eq)}(\tilde{\omega})
= - \pi^{-1} {\rm Im} [\hat{g}^{R}_{{\bf k}\xi\alpha}(\omega)]_{bb;nn}$ with Eq.~(\ref{TunnelingFormula_3}) reducing to the well-known 2D-to-2D tunneling current formula \cite{2D_Tunneling1,2D_Tunneling2}. 
% In this subsection, we reveal the underlying mechanism of resonant photon-assisted tunneling. For simplicity, we use Eq.~\eqref{TunnelingFormula_3} instead of Eq.~\eqref{TunnelingFormula_2}, ignoring the vertex correction to the tunneling Hamiltonian. The effect of the vertex correction will be discussed in the next subsection. 
% Since essential informations of Eq.~\eqref{TunnelingFormula_3} are encoded in the photon-dressed spectral function [Eq.~\eqref{PhotonDressedSpectralFunction}], we first try to catch its property. 
Fig.~\ref{Figure1}(a) shows the Floquet mode  %photon-dressed 
spectral function $\rho_{{\bf k}\xi\alpha b}^{\rm (ph)}(\tilde{\omega})$ 
%Eq.~\eqref{PhotonDressedSpectralFunction} 
%for valley $K'$, layer $\alpha\in\{T,B\}$, and sublattice $b$ 
as a function of the equilibrium energy dispersion $\mathcal{E}_{\bf k}$ for different values of the driving field $E$. At small values of $E$, the quasienergy dispersions of the Floquet sidebands (\textit{i.e.}, photon-dressed electronic bands)
%which are electronic bands dressed by photons) 
approximately coincide with copies of the equilibrium bands shifted by integer multiples of $\hbar \Omega$, and the reduced-zone Floquet quasienergy $\bar{\mathcal{E}}_{\bf k} \approx \mathcal{E}_{\bf k}$. The undressed conduction and valence bands each generate their own Floquet sidebands, which for convenience will be called Floquet conduction  and valence sidebands (FCSB and FVSB, respectively). %The spectral weight decreases as the order $n$ of the Floquet sideband increases. 
%Higher order Floquet sidebands 
%Their spectral weight of the Floquet states decreases as $n$ increases away from $n = 0$. 
%Associated with each of the 
%Since there are conduction and valence bands in equilibrium, under illumination each of these bands will produce a series of ``Floquet copies'' that are primarily conduction %and valence in character, which can be called Floquet conduction and valence sidebands. 
%For small $E$, regarding the $n=0$ Floquet sideband belonging to $\gamma=-1$, we find that the numerical results and the guideline (red dotted line), thus indicating %$\bar{\mathcal{E}}_{\bf k} \approx \mathcal{E}_{\bf k}$. 
As $E$ increases, mixing between Floquet states becomes stronger and quasienergy band gaps appear prominently at the locations of anti-crossing, given by $(\mathcal{E}_{\bf k}, \hbar\tilde{\omega} - \mu_\alpha)$ $= (m-n, m+n) \hbar\Omega/2$  with $m,n\in\mathbb{Z}$. 
%For larger $E$, the anti-crossing becomes dominant so that our assumption holds no longer. 

%%%%%%%%%%%%%%%%%%%%%%%%%%%
%%%%%%%%%%%%%%%%%%%%%%%%%%%

\begin{figure}[t]
\centering
\includegraphics[width=0.48\textwidth]
{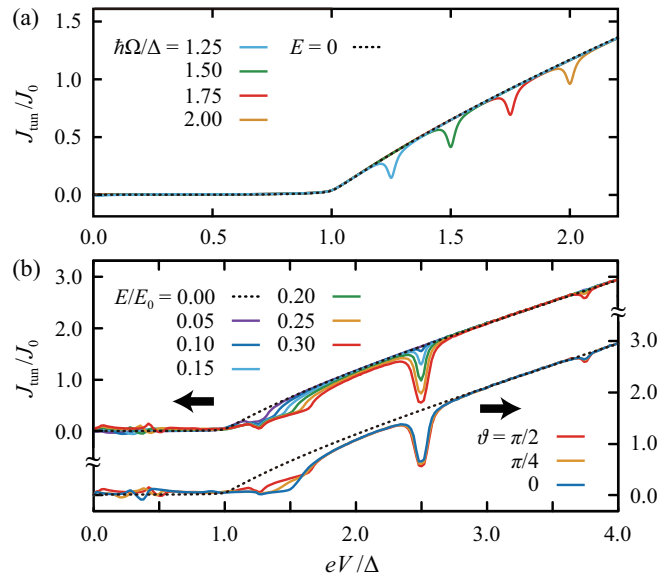} \\
\caption{
Photon-assisted tunneling current density $J_{\rm tun}$ in units of $J_0$ as a function of bias voltage $V$.
(a) Solid lines show the numerical results obtained from the second-order perturbation theory for a  fixed $E/E_0 = 0.015$ and different values of $\Omega$. 
The dashed line shows the dark tunneling current. 
(b) Exact numerical results [Eqs.~\eqref{TunnelingFormula_2}-\eqref{PhotonDressedInterlayerTunnelingHamiltonian}] for $\hbar\Omega/\Delta = 1.25$. The upper set of solid lines shows the tunneling current at a fixed polarization angle $\vartheta = \pi/2$ for different values of the optical field strength $E$. The lower set of solid lines shows the tunneling current at a fixed $E/E_0 = 0.3$ for different values of polarization angle $\vartheta$. 
In all figures, we use the common parameters $k_B T / \Delta = 0.02$, $\Gamma / \Delta = 0.01$, and $\mathcal{E}_c / \Delta = 4$~\cite{Comment3}.
}
\label{Figure2}
\end{figure}

\textit{Single-photon excitations} --- Before carrying out fully numerical calculations of Eq.~(\ref{TunnelingFormula_2}), we first perform a second-order perturbative analysis in the driving field amplitude. In this work, we focus on low temperatures $k_B T \ll \Delta$ and evaluate the tunneling current, 
%in the zero-temperature limit, 
assuming each graphene layer is at half-filling so that the electrochemical potential under bias is $\mu_T = - \mu_B = eV/2$. Treating the light field $\mathcal{V}_\xi(t)$ as a perturbation and expanding the single-layer Green's function into $\hat{\mathcal{G}}_{{\bf k}\xi\alpha}(\omega) = \sum_{j=0,1,2} (\mathcal{V}_0)^{j} \hat{\mathcal{G}}_{{\bf k}\xi\alpha}^{(j)}(\omega) + \mathcal{O}(\mathcal{V}_0^3)$~\cite{Comment1}, Eq.~(\ref{TunnelingFormula_2}) can be written in the form  $J_{\rm tun}  = \sum_{j=0,1,2} (\mathcal{V}_0)^{j} J_{\rm tun}^{(j)} + \mathcal{O}(\mathcal{V}_0^3)$, where the zeroth order term \cite{J_0_expression} yields the dark tunneling current \cite{Vasko13,Brey,Polini}. 
%
% \begin{align}
% \frac{J_{\rm tun}^{(0)}}{J_0}
% & = \frac{\pi}{4} \bigg( \frac{eV}{\Delta} - \frac{\Delta}{eV} \bigg) \big[ \Theta(eV - \Delta) - \Theta(eV - 2\mathcal{E}_c) 
% \nonumber\\
% & ~~~ + \Theta(eV + 2\mathcal{E}_c) - \Theta(eV + \Delta) \big],
% \label{J_0_a}
% \end{align}
% %
% where $\Theta(\cdot)$ is the step function. 
%in agreement with Ref.~\cite{Vasko13}. 
The first-order contribution $J_{\rm tun}^{(1)} = 0$ because $J_{\rm tun}^{(1)}(t) \propto \sin(\Omega t)$ vanishes under time-averaging. 
Fig.~\ref{Figure2}(a) shows, for different values of frequency, the tunneling current $J_{\rm tun}  =  J_{\rm tun}^{(0)}+\mathcal{V}_0^{2} J_{\rm tun}^{(2)}$ obtained within the present second-order perturbation theory. It is seen that the overall profile remains close to the dark tunneling current $ J_{\rm tun}^{(0)}$ (dashed line) except at $eV = \hbar\Omega$. When the bias voltage is equal to the photon energy, tunneling transitions assisted by a single photon occur, manifesting as a single anti-resonance in the second-order contribution $J_{\rm tun}^{(2)}$.

\textit{Photon-assisted tunneling suppression} --- 
To account for strong field and multiphoton excitation effects,  we now evaluate the tunneling current from Eqs.~\eqref{TunnelingFormula_2}-\eqref{PhotonDressedInterlayerTunnelingHamiltonian} non-perturbatively. Fig.~\ref{Figure2}(b) shows our full numerical results 
%calculations of the tunneling current from Eqs.~\eqref{TunnelingFormula_2}-\eqref{PhotonDressedInterlayerTunnelingHamiltonian} in  Fig.~\ref{Figure2}(c)   
% Further features arise if we consider full numerical calculations from Eqs.~\eqref{TunnelingFormula_2}-\eqref{PhotonDressedInterlayerTunnelingHamiltonian}, which describe the non-perturbative light coupling effect. Figure~\ref{Figure2}(c) shows the counterpart of (a)
for different field strengths $E$ (upper set of solid lines) and for different polarization angles $\vartheta$ (lower set) with an optical frequency value $\hbar\Omega = 1.25\Delta$ slightly larger than the band gap. 
%fixed value of optical frequency $\Omega$. %Overall features are consistent with Fig.~\ref{Figure2}(a). 
In the absence of light, the dark tunneling current %Eq.~(\ref{J_0_a}) 
remains zero when the bias voltage is smaller than the band gap, as shown by the dashed lines in both Figs.~\ref{Figure2}(a)-(b). 
When light is turned on, we first notice that the PAT current becomes non-zero even when $eV < \Delta$, a feature not captured by the second-order perturbation theory. 
% and highlights the importance of the non-perturbative effects of illumination. 
For bias values greater than the band gap $eV > \Delta$, our theory predicts periodic anti-resonance suppression of the tunneling current with a separation $\hbar\Omega$ along the bias voltage axis as seen in Fig.~\ref{Figure2}(b).
%is dramatically suppressed to almost zero from its dark value due to the first anti-resonance located at $eV = \hbar\Omega$.  
%For bias values greater than but close to the band gap $eV \gtrsim \Delta$, we find a considerable  suppression in the PAT current as compared to the dark tunneling current. 
%As $E$ increases, we observe an enhancement in the tunneling current for $eV < \Delta$ and an overall suppression for $eV > \Delta$ on top of the dark tunneling current. 
%For even larger values of $eV > \Delta$, our theory predicts periodic anti-resonance suppression of the tunneling current with a separation $\hbar\Omega$ along the bias voltage axis. 
%at quantized values of the photon energy along the bias voltage axis. 
%that are periodically developed with a separation $\hbar\Omega$ along the bias voltage axis. 
%Remarkably, we find that 
Remarkably, the anti-resonance positions are found to be independent of %the values of 
the field strength $E$ and polarization angle $\vartheta$, and are precisely given by integer multiples of $\hbar\Omega$, \textit{i.e.}, $eV = N\hbar\Omega$ with $N \in \mathbb{Z}$. %The strength of the peak is strongest when $eV = \hbar\Omega$ and decreases with the order $N$ of the anti-resonance. 
%While our previous second-order perturbation analysis can capture the first-order ($n = 1$) anti-resonance, it is limited within the weak field regime $\mathcal{V}_0 \ll \Delta$ %and does not predict the appearance of higher order ($n \geq 2$) anti-resonances. 
Close to the band gap $eV \gtrsim \Delta$ in particular, we find a dramatic suppression of the PAT current to almost zero from its dark value due to the first anti-resonance located at $eV = \hbar\Omega = 1.25\Delta$.  
Fig.~\ref{Figure2}(b) also shows that the strength of the anti-resonances increases with the field strength,  
%of the optical field $\mathcal{V}_0$, 
resulting in a progressive suppression of the tunneling current and indicating a tendency towards complete localization at stronger fields.  
%Up to $E = 0.3E_0$, the tunneling suppression at $eV = \hbar\Omega$ is found to be quite substantial, nearly reaching complete suppression. This suggests the possibility of a %complete %suppression of interlayer tunneling (\textit{i.e.}, localization) 
%localization when the optical field strength is increased further. %, which will be the subject of a future investigation. 
% resulting in a strong suppression of the tunneling current. Up to $E = 0.3E_0$, the tunneling suppression at $eV = \hbar\Omega$ is found to be quite substantial, nearly reaching complete suppression. This suggests the possibility of a complete %suppression of interlayer tunneling (\textit{i.e.}, localization) 
% localization when the optical field strength is increased further. %, which will be the subject of a future investigation. 
%Here, we emphasize that 
%Here we note 
Note here that the nature of our predicted 
tunneling suppression under optical illumination 
is fundamentally different from dynamic localization \cite{DL,Grifoni98,Kohler}, which occurs only when the coupling parameter $\mathcal{V}_0/\hbar\Omega$ is equal to a zero of the Bessel function. 
%The latter case occurs under a different condition when the coupling parameter $\mathcal{V}_0/\hbar\Omega$ is equal to a zero of the Bessel function. 
%; whereas in our case the tunneling suppression occurs precisely at integer multiples of $\hbar\Omega$. 

%%%%%%%%%%%%%%%%%%%%%%%%%%%
\begin{figure}[t]
\centering
\includegraphics[width=0.48\textwidth]
{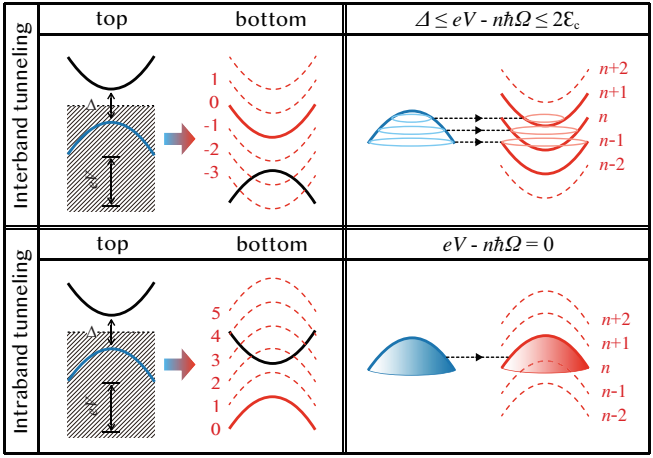} \\
\caption{
Schematic for forward tunneling transitions for $eV>0$. 
The left panel depicts the interband and intraband tunneling mechanisms assisted by multiphoton processes. 
%Only the forward tunneling contribution is shown, the backward tunneling process is similar. %\cite{Comment1}. 
%for the two main mechanisms of photon-assisted tunneling, including interband and intraband tunnelings for $V>0$. Here, 
The Floquet sidebands are indexed by red-colored integers. 
The right panel depicts the resonance tunneling conditions for the interband (or intraband) tunneling processes 
%for the corresponding types of mechanisms,  
%. Here, we focus on the 
%showing a forward tunneling process 
from %an original 
the undressed valence band of the top layer to a Floquet conduction (or valence) sideband of the bottom layer. Backward tunneling transitions can be visualized  similarly from  Eq.~(\ref{TunnelingFormula_3}). 
}
\label{Figure3}
\end{figure}
%%%%%%%%%%%%%%%%%%%%%%%%%%%

The periodic occurrence of the tunneling suppression uncovered by our calculations stems from resonant intraband tunneling assisted by multiphoton excitations.
%as photon-assisted resonant tunneling. 
%First, it will be useful to 
Let us first consider 
%First we consider 
the case without illumination. Due to conservation of in-plane momentum, conventional 2D-to-2D intraband tunneling in coupled semiconductor quantum wells occurs when the layers' Fermi levels are closely matched, and resonance tunneling happens at exact matching whereby $J_{\mathrm{tun}} \sim \tau$ becomes sharply peaked (here $\tau$ is the electron's lifetime) ~\cite{Eisenstein}. %This tunneling process is an intraband transition by nature. 
%The tunneling current $J_{\mathrm{tun}} \sim 1/\tau$ (where $\tau$ here is the electron's lifetime) when the layers' Fermi levels are mismatched is indicative of a dissipative response while $J_{\mathrm{tun}} \sim \tau$ at resonance~\cite{Eisenstein} is indicative of a coherent response. 
%When the layers' Fermi levels are mismatched but within the width of electron's lifetime broadening, the tunneling current $J_{\mathrm{tun}} \sim 1/\tau$ (where $\tau$ here is the electron's lifetime) is dissipative
%when the layers' Fermi levels are mismatched is indicative of a dissipative response while $J_{\mathrm{tun}} \sim \tau$ at resonance~\cite{Eisenstein} is indicative of a coherent response.
%The tunneling current $J_{\mathrm{tun}} \sim 1/\tau$ when the layers' Fermi levels are mismatched is indicative of a dissipative response while $J_{\mathrm{tun}} \sim \tau$ at resonance~\cite{Eisenstein} is indicative of a coherent response. 
%Semimetals or semiconductors, including graphene in our case, 
%also 
Graphene allows for interband tunneling when the bias voltage exceeds the band gap. 
%The interband tunneling mechanism is unique to semimetals or small-gap semiconductors, including graphene in our case. 
If illumination is absent in our system, only interband (but not intraband) tunneling can occur 
%there can only be interband tunneling but no intraband tunneling, 
since the layers' Fermi levels are assumed to be inside the band gap. However, an optical driving field opens up many additional channels for intraband tunneling via photon-assisted transitions between Floquet sidebands (see Fig.~\ref{Figure3}). %, which are electronic bands dressed by photons. 
%Since there are conduction and valence bands in equilibrium, under illumination each of these bands will produce a series of ``Floquet copies'' that are primarily conduction %and valence in character, which can be called Floquet conduction and valence sidebands. 
When one layer's undressed valence band (VB) edge is aligned with one of the many FVSB edges of the other layer, electron momentum and energy conservation are simultaneously satisfied and resonant tunneling can occur. This Floquet band edge alignment happens when the bias voltage is tuned to an integer multiple of the driving frequency. 
%, $eV = N\hbar\Omega$. 
To understand why a suppression instead of an enhancement occurs, it is useful to refer to Eq.~(\ref{TunnelingFormula_3}) under a forward bias condition $eV > 0$. 
% Without illumination and under a forward bias condition $eV > 0$, the forward tunneling contribution (\textit{i.e.}, first term) is dominant. When light is turned on under the resonant condition $eV = n\hbar\Omega$, we find that the backward tunneling contribution (second term) is dramatically enhanced, strongly suppressing the forward tunneling current.  
Physically, the forward tunneling contribution involves tunneling from the undressed VB in the \textit{top} layer to the $n > 0$ FVSBs in the \textit{bottom} layer (Fig.~\ref{Figure3}), while the backward tunneling contribution involves tunneling from the undressed VB in the \textit{bottom} layer to the $n < 0$ FVSBs in the \textit{top} layer. 
%Physically, the backward tunneling contribution involves tunneling from the undressed valence band in the \textit{bottom} layer to the $n < 0$ Floquet valence bands in the %\textit{top} layer, while the forward tunneling contribution involves tunneling from the undressed valence band in the \textit{top} layer to the $n > 0$ Floquet valence bands in %the \textit{bottom} layer (see Fig.~\ref{Figure3}). 
%Compared to the $n > 0$ Floquet valence sidebands, the $n < 0$  Floquet valence sidebands carry a higher spectral weight, as confirmed by our numerical calculations of integrated spectral weights of the photon-dressed spectral function [Fig.~\ref{Figure1}(b)]. To zeroth order in the optical field strength, the Floquet quasienergy spectrum consists of identical copies of the undressed conduction band and valence band displaced by integer multiples of $\hbar\Omega$. 
% Field-induced mixing will be stronger between the $n > 0$  Floquet valence sidebands and the $n < 0$ Floquet conduction sidebands
% Field-induced mixing between the $n < 0$ Floquet conduction sidebands
% because they are more separated in quasienergy from the $n < 0$ Floquet conduction sidebands
% , with which 
%
% field-induced mixing between the $n < 0$  Floquet valence sidebands and the $n < 0$ Floquet conduction sidebands is weaker because they are more separated in quasienergy. Therefore, the $n < 0$  Floquet valence sidebands carry a higher spectral weight than their $n > 0$ counterparts, 
%
%Due to its reduced band mixing with $n < 0$ FCSBs, 
Because they are further removed from the $n < 0$ FCSBs, the $n < 0$ FVSBs carry a higher spectral weight than their $n > 0$ counterparts \cite{specfun}, 
as confirmed by our numerical calculations of integrated spectral weights of the Floquet mode spectral function [Fig.~\ref{Figure1}(b)].  
As a result, under the resonant condition $eV = N\hbar\Omega$, the backward tunneling contribution is dramatically enhanced causing a suppression of $J_{\rm tun}$. 
\begin{figure}[t]
\centering
\includegraphics[width=0.48\textwidth]
{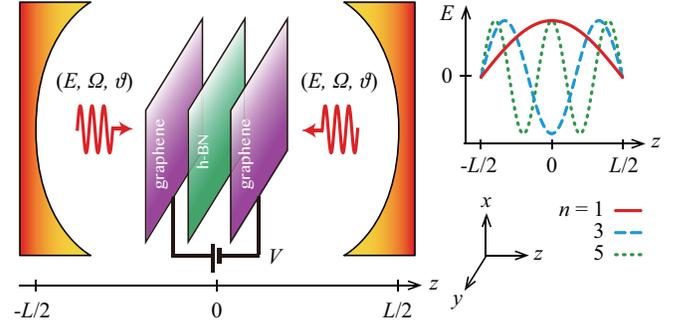} \\
\caption{
Schematic for experimental proposal. Graphene/{\it h}-BN/Graphene heterostructure is positioned at the center of the optical cavity (left), in which an odd cavity mode is established (right).
 %The optical field inside the cavity is calibrated to form standing waves with odd modes. 
}
\label{Figure4}
\end{figure}
%%%%%%%%%%%%%%%%%%%%%%%%%%%

\textit{Proposed experimental setup} --- We close by commenting on the conditions for observability. The assumption that the electromagnetic field on the two graphene layers are the same is not a stringent requirement and using a single light source to illuminate the system does not preclude the resonant tunneling suppression from occurring. Taking account of light reflections, the graphene layer further removed from the light source will experience a driving field with a weaker amplitude but nonetheless the same frequency. The resonant tunneling suppression is therefore expected to occur at the same values of bias voltage $eV = N\hbar\Omega$ albeit with less pronounced anti-resonances. To realize the condition with the same electric field ampitudes on both layers using a single laser, two alternative scenarios can be devised. First, this condition can be approximately achieved when the laser wavelength is long compared with the thickness of the vdWH.
%illuminating the vdWH using a laser having a wavelength that is long compared with than thickness of the vdWH. %($\sim \mathrm{nm}$). 
Second, an optical cavity can be used (see Fig.~\ref{Figure4}). When the cavity mode of the standing wave is odd and the vdWH is placed at the cavity's center, both graphene layers will experience the same magnitude of electromagnetic field. 

The resonant intraband tunneling mechanism we discovered is a generalization of the usual dark resonant tunneling to the scenario with Floquet sidebands under strong optical illumination. %We mention that 
This phenomenon should apply not only to graphene layers, but also to trilayer vdWHs with other 2D materials such as bilayer graphene and transition-metal dichalcogenides. The fact that the tunneling current can be turned on and off by illumination at frequency values equal to an integral fraction of the bias voltage suggests a time-dependent control scheme for switching applications, opening the door to dynamical tuning of tunneling dynamics using periodic drives.   

\acknowledgments
%%%%%%%%%%%%%%%%%%%%%%%%%%%

We thank Patrick Kung, Takashi Oka, and Godfrey Gumbs for useful discussions. This work was supported by startup funds from the University of Alabama and the U.S. Department of Energy, Office of Science, Basic Energy Sciences under Early Career Award $\#$DE-SC0019326.

%We gratefully acknowledge support by startup funds from the University of Alabama and the U.S. Department of Energy Early Career Award $\#$DE-SC0019326.

%This work was supported by startup funds from the University of Alabama and the U.S. Department of Energy Early Career Award $\#$.

%. WKT gratefully acknowledges support from the U.S. Department of Energy, Office of Science, Basic Energy Sciences under Award #

\newpage

%%%%%%%%%%%%%%%%%%%%%%%%%%%


\begin{thebibliography}{99}
\bibitem{J_0_expression} A closed-form analytic expression can be obtained, see \cite{Comment1}.
\bibitem{specfun} %For a two-band system, it is useful to 
If one imagines the formation of Floquet sidebands in two steps: (1) generation of ``Floquet copies'' via displacement of the  undressed conduction and valence bands each by $n\hbar\Omega$ (where $\vert n \vert = 0, 1, 2, \dots$) and (2) band-mixing between the Floquet copies, then one can readily understand that the $n < 0$  FVSBs have a higher spectral weight  compared to the $n > 0$ FVSBs, because the $n < 0$ Floquet copies of the valence band and  $n > 0$ Floquet copies of the conduction band have less overlap  and hence less band-mixing that would result in spectral gaps in the quasienergy dispersions.
\end{thebibliography}
\end{document}